# Direct Modulation of Electrically Pumped Coupled Microring Lasers

CHI XU,[1] WILLIAM E. HAYENGA[1], DEMETRIOS N. CHRISTODOULIDES[1], MERCEDEH KHAJAVIKHAN[2,3], AND PATRICK LIKAMWA[1,4]

[1]CREOL, The College of Optics and Photonics, University of Central Florida, Orlando, Florida 32816, USA
[2]Ming Hsieh Department of Electrical and Computer Engineering, University of Southern California, CA 90007, USA
[3]khajavik@usc.edu,
[4]patrick@creol.ucf.edu

**Abstract:** We demonstrate how the presence of gain-loss contrast between two coupled identical resonators can be used as a new degree of freedom to enhance the modulation frequency response of laser diodes. An electrically pumped microring laser system with a bending radius of 50 μm is fabricated on an InAlGaAs/InP MQW p-i-n structure. The room temperature continuous wave (CW) laser threshold current of the device is 27 mA. By adjusting the ratio between the injection current levels in the two coupled microrings, our experimental results clearly show a bandwidth improvement by up to 1.63 times the fundamental resonant frequency of the individual device. This matches well with our rate equation simulation model.

## 1. Introduction

Directly modulated semiconductor laser diodes have several attractive features such as low cost and potential for high-density integration [1, 2]. However, the intrinsic modulation bandwidth of a laser diode tends to be capped by its relaxation resonance frequency, leading to limited applications in optical communication systems [3]. Over the years, a number of technologies have been developed for realizing high-speed directly modulated semiconductor laser diodes. These include but are not limit to, nanolasers with strong Purcell effect [4-7], narrow linewidth quantum dot lasers [8], optical injection locking [9, 10], and photon-photon resonance in coupled vertical-cavity surface-emitting lasers (VCSELs) and edge emitting lasers [11-14].

Microring resonators are one of the most attractive cavity structures in photonic integrated circuits. Due to the absence of reflective facets, ring resonators can enable lasers with high quality factors, that are compact for on-chip device integration. Several efforts have also been devoted to improving the performance of semiconductor ring lasers in practical applications, such as lowering their threshold [15], unidirectional emission [16], and compatibility with silicon [17]. In recent years, non-Hermiticity has been widely used in photonic settings in order to delicately mold light-matter interactions, realize sensitivity enhancements [18-21], enforce unidirectional invisibility [22, 23], and enable topological lasers [24, 25]. One widely used approach for establishing non-Hermiticity is to use two coupled cavities one subject to gain and the other to loss (or a less amount of gain). For example, a laser consisting of a coupled ring resonator systems with differential gain have been demonstrated to enhance the mode suppression ratio and promote single mode lasing [26-28]. In this work, we investigate the use of gain and loss in a dual ring laser system as a new mechanism to modify the maximum modulation speed of the laser diode.

In this paper, we explore the interplay between unevenness of pumping and modulation bandwidth in the coupled microring configuration. The paper is structured as follows: In

Section 2, we study the modulation characteristics of this coupled laser system using the rate equation model. In Section 3, we provide the electromagnetic mode simulations for a laser designed on a InAlGaAs/InP multiple quantum well (MQW) epitaxial wafer. In Section 4, we describe the steps involved in the fabrication of the electrically pumped microring lasers. In Section 5, we characterize the laser and measure the frequency responses under different pumping ratios. Finally, Section 6 concludes the paper.

## 2. Rate equation model for coupled ring laser system

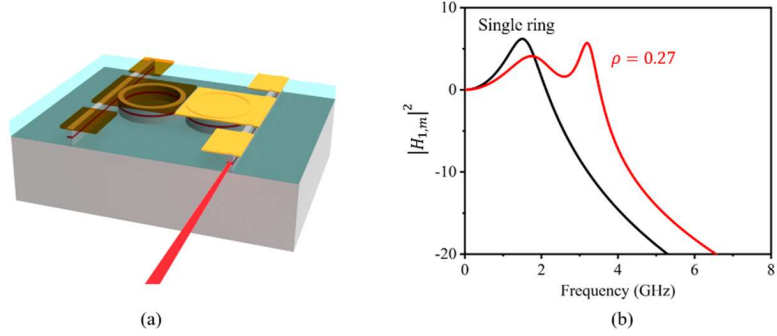

(a)

(b)

**Fig.1.** (a) The schematic diagram of an electrically pumped coupled microring laser system; (b) simulated frequency response of coupled microring lasers with $\rho = 0.27$.

Figure 1 (a) shows a schematic of the coupled ring laser system. It consists of two identical microring cavities that are evanescently coupled to each other due to proximity. Photon resonance can be supported in this coupled cavity system by applying a sufficient amount of pump. Consequently, the laser emission is extracted through a coupled bus waveguide positioned in the side of one of the rings. Each cavity has its own electrodes that allows a diverse set of pumping schemes. Considering only the fundamental mode $TE_{10}$ is supported, the electric field of $TE_{10}$ in each cavity $E_1, E_2$ can be described using the following coupled mode equations in the temporal domain:

$$\dot{E_1} = (G - \gamma)(1 - j\alpha)E_1 + j\kappa E_2, \tag{1.a}$$

$$\dot{E_2} = -(F + \gamma)(1 - j\alpha)E_2 + j\kappa E_1, \tag{1.b}$$

where $\gamma$ signifies the linear loss of a passive resonator mainly due to scattering, bending and output coupling losses, $G$ and $F$ stand for carrier induced gain and loss in the respective rings. One thing to be noticed, F can be the loss or a lower level of gain depending on the carrier density. $\alpha$ represents the linewidth enhancement factor, and $\kappa$ is the temporal coupling coefficient [26].

Equation 1 depicts the dynamics of photons in two rings. Considering an electrical pumping scheme, the laser rate equation of this coupled structure can be described by:

$$\dot{N_1} = \eta_I I_1/(qV) - N_1/\tau_N - \sum v_g g_m S_{1,m}, \tag{2.a}$$

$$\dot{S_{1,m}} = \Gamma v_g g_m S_{1,m} - S_{1,m}/\tau_p - 2\kappa sin\phi_m \sqrt{S_1 S_2}, \tag{2.b}$$

$$\dot{\phi_m} = \alpha\Gamma v_g(g_m + f_m)/2 + \kappa cos\phi_m(1/\rho - \rho), \tag{2.c}$$

$$\dot{S_{2,m}} = -\Gamma v_g f_m S_{2,m} - S_{2,m}/\tau_p + 2\kappa sin\phi_m \sqrt{S_{1,m} S_{2,m}}, \tag{2.d}$$

$$\dot{N_2} = \eta_I I_2/(qV) - N_2/\tau_N + \sum v_g f_m S_{2,m}. \tag{2.e}$$

where $I_i$ is the injected current of ring $i$ ($i = 1,2$), $q$ is the elementary charge, $V$ is the volume of the active region, $\eta_I$ stands for the current injection efficiency, and $S_{i,m}$ and $N_i$ are the carrier density and photon density of the $m^{th}$ mode in ring $i$. In these equations, mutually coupled modes are placed in the same order. The cavities are designed to support single transverse mode

TE$_{10}$. Therefore, a pair of counter-propagating modes in adjacent rings share the same $m$. In these equations, $\rho$ and $\phi$ stand for the ratio of modal field amplitudes and phase difference, respectively, $\tau_N$ is the carrier lifetime, $g$ and $f$ represent gain and loss related to the temporal counterpart $G$ and $F$ via $2G = \Gamma v_g g$ and $2F = \Gamma v_g f$, respectively, $\tau_p$ is the photon lifetime, $v_g$ is the group velocity, and $\Gamma$ is the confinement factor. In the following discussion, $\tau_p$, $v_g$ and $\Gamma$ are assumed to be the same for the two rings due to their identitical structures.

At steady state, phase difference and modal field ratio can be determined by gain-loss contrast $\delta = G + F$ via $\phi = \pi + \arctan\left[\alpha^{-1}(\rho^2 - 1)/(\rho^2 + 1)\right]$ and $\delta = \kappa(\rho + 1/\rho)$. The modulation frequency response of the $m$th order mode outcoupling from the gain cavity (ring 1), i.e. $|H_{1,m}| = |\Delta S_{1,m}/\Delta I_1|$, can be obtained by applying small sinusoidal signal to its steady state solution:

$$(j\omega\mathbf{I} - \mathbf{J})\Delta X = \eta_l(qV)^{-1}\Delta Y \tag{3}$$

where $X = (N_1, S_1, \phi, S_2, N_2)$, and $Y = (I_1, 0, 0, 0, 0)$ can be expressed as $X = X_0 + \Delta X e^{j\omega t}$, $Y = Y_0 + \Delta Y e^{j\omega t}$ under the modulation. Here, $\mathbf{J}$ is the Jacobian of Eq. (2). From Eq. (2) and Eq. (3), $|H_{1,m}|$ not only depends on $S_{1,m}$ and $\tau_p$, but is also subject to $\delta$.

**Table 1. Parameters of InAlGaAs MQW microring laser**

| Symbol | Unit | Value |
|---|---|---|
| $V$ | cm$^3$ | $4.14 \times 10^{-11}$ |
| $\Gamma$ | 1 | 0.01 |
| $\eta_l$ | 1 | 0.4 |
| $\tau_p$ | ps | 26 |
| $I_{th}$ | mA | 25 |
| $a$ | cm$^2$ | $1.6 \times 10^{-15}$ |
| $\lambda$ | nm | 1550 |
| $v_g$ | cm/s | $9.4 \times 10^9$ |

To verify how the gain-loss contrast can modify the features of $H(j\omega)$, we numerically simulate the modulation response by employing the rate equation model to a InAlGaAs MQW laser, with the parameters given in Table 1. Figure 1(b) contrasts the modulation response of a single ring laser to that of a coupled configuration with $\kappa = 5$ GHz and $\rho = 0.27$. The single cavity laser exhibits a bandwidth of 2.3 GHz at an excess pumping of $I - I_{th} = 8$ mA. By increasing the interaction between the two cavities ($\rho = 0.27$), the modulation bandwidth boosts to 3.7 GHz. This response can be attributed to the shortening of effective photon lifetime $1/\tau_p' = 1/\tau_p + 2\kappa\rho\sin\phi$. A comperehansive analysis of this behavior can be found in our previous work [29].

## 3. Design and mode analysis

In this study, we implemted our lasers on an InAlGaAs MQW gain medium. These quantum well structures tend to have a more favorable thermal performanc due to their larger conduction band offset ($\Delta E_c = 0.72\Delta E_g$) [30]. The structure of InAlGaAs MQW epitaxial wafer is shown in **Error! Reference source not found.**. The epitaxial layers were grown on a n-doped InP substrate using metal organic chemical vapor deposition (MOCVD). The intrinsic layers consist of five-period of MQW with $\lambda_{PL} = 1508$ nm, sandwiched between two 100 nm thick separate confinement heterostructure (SCH) layers. The p-type layers consist of a heavily doped InGaAs cap for metallization, followed by a 1.615 μm Zn-doped InP cladding and a 100 nm InAlAs etch-stop layer. The lower n-doped layers are comprised of a 140 nm thick InAlAs, a 500 nm thick InP buffer layer, and the substrate.

**Table 2. InAlGaAs MQW wafer structure**

| | |
|---|---|
| P-InGaAs Layer | 150 nm |
| P-1.5Q Layer | 25 nm |
| P-1.3Q Layer | 25 nm |
| P-InP | 1500 nm |
| P-(1.1Q~1.15Q) Layer | 15 nm |
| P-InP | 50 nm |
| U-In$_{0.52}$Al$_{0.48}$As | 50 nm |
| U-GRIN-In$_{0.53}$Al$_x$Ga$_{0.47-x}$As | 100 nm |
| QW/Barrier ($\lambda_{PL}$) | 5-6/9 (1508) nm |
| U-GRIN-In$_{0.53}$Al$_x$Ga$_{0.47-x}$As | 100 nm |
| N-In$_{0.52}$Al$_{0.48}$As Layer | 140 nm |
| N-InP Buffer Layer | 500 nm |
| InP Substrate | |

In designing a microring cavity with a low bending loss and a single transverse mode, the finite element method (FEM) simulation module FemSIM of a commercial photonic simulation software Rsoft is used. The cross-section of the fundamental mode profile $|E_t|$ is shown in Fig.2(b), exhibiting a shift towards the outer sidewall due to bending. The waveguide is 1.65 μm wide with a bending radius of 50 μm. It is deeply etched through quantum wells for a high quality factor Q that is determined by $Q = n_r/2n_i = 2.8 \times 10^5$, where $n_r$ and $n_i$ are the real and imaginary parts of mode's effective index, respectively.

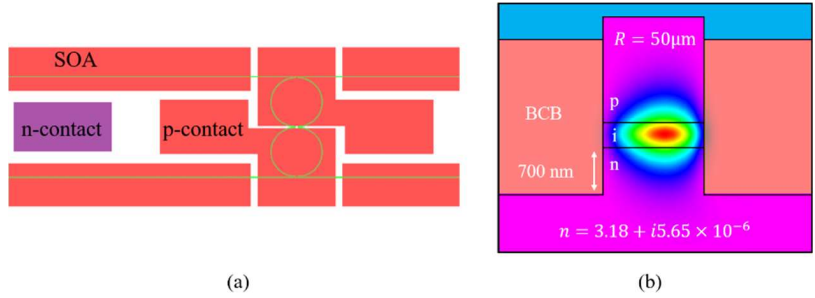

(a)                                    (b)

**Fig.2.** (a) The top view layout of designed microring lasers; (b) the cross-section mode profile of microring cavity.

In our design, the coupling between the two ring resonators is achieved based on directional coupling, which comprises a straight waveguides section with a gap of $g$. Coupling in the bending region can be ignore due to the strong confinement and rapid deviation of two modes. The temporal coupling coefficient is related to that in the spatial domain through $\kappa = v_g \kappa' l_c/(2\pi R)$, where $l_c$ represents the coupling length. The mode profiles of the odd and even supermodes in the cross-section are shown in Fig.3. The spatial coupling coefficient determined by the effective index of two supermodes, is $2.7 \times 10^3$ m$^{-1}$. The temporal coupling coefficient is 4.9 GHz if $l_c$ is chosen as 6 μm. In practice, the imprefections in fabrication such as sidewall roughness or etching ripples tend to increase the coupling [28].

Light is extracted from straight bus waveguides which serve as semiconductor optical amplifiers (SOA) to boost the output signal. Due to the spiral symmetry of the ring cavity, two counter-propagating modes' encounters are same. Therefore, one can collect mutually coupled modes from either side of the chip.

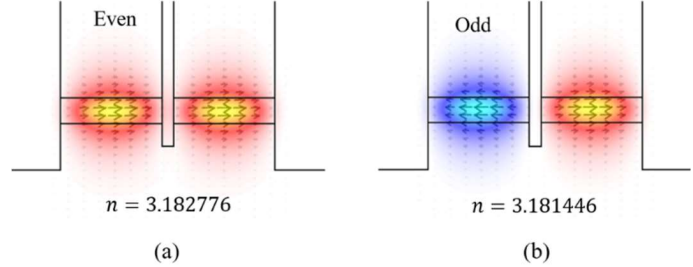

| Even | Odd |
|:---:|:---:|
| $n = 3.182776$ | $n = 3.181446$ |
| (a) | (b) |

**Fig.3.** Cross-section mode profiles of even (a) and odd (b) modes of a directional coupler.

## 4. Fabrication procedures

Figure 4 shows the processing steps involved in the fabrication of the electrically pumped microring lasers. A cleaned sample was first immersed into a 3:1 mixture of HCl:$H_3PO_4$ to remove the 150 nm thick InP protective layer. E-beam lithography is then followed by an inductively coupled plasma – reactive ion etching ptocess (ICP-RIE) with $CH_4$/$H_2$/$Cl_2$ (3/7/8 SCCM) in order to transfer the device layout to the epi-wafer. Since an extended dry etching process needed to etch through the MQW region, it will inevitably increase the surface defect density on the walls of the gain medium leading to a high threshold current as a result of strong non-radiative recombination. Consequently, a passivation treatment of the etched wall is required [31]. This is accomplished by first immersing the sample into a solution of $H_3PO_4$/$H_2O_2$/$H_2O$ (1/1/38) for 6 s, followed by soaking in a solution of 20% $(NH_4)_2$S that is further diluted in $H_2O$ in the ratio of 1:10 for 5 minutes to remove the surface defects and form protective monolayers. Then the sample was immediately dried with $N_2$ gas without any rinsing, and coated with a 100 nm thick layer of $SiO_2$. Next, the top n-contact was first created by depositing Ni/Ge/Au (7/20/200 nm). After benzocyclobutene (BCB) planarization, the p-type metal contact metals Ti/Zn/Au (7/4/500 nm) were evaporated on the sample, followed by a rapid thermal annealing (RTA) at 400C° for 1 min.

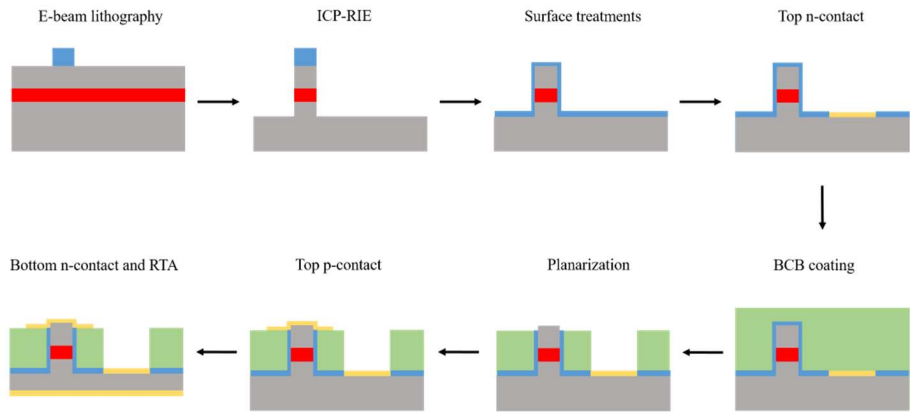

**Fig.4.** Fabrication processes involved in realizing the electrically pumped microring lasers.

## 5. Laser characterization and modulation response

The schematic of the test setup for measuring the frequency response is shown in Fig. 5. The probed laser placed on a heat sink is assembled in a micro-positioner. By cleaving the sample, light emission is collected from the facet by a 20X objective lens. A removable mirror is used to direct the laser emission to either an infrared camera or a high-speed photodetector (Newport 818-BB-35). For the alignment, the cleaved facet is illuminated by a 1310 nm laser diode. DC bias current from a laser diode driver (LDX-3500) and RF signal current from Port 1 of the vector network analyzer (Aglient 8720E) are combined by a bias tee, then injected into the laser through a picoprobe. Direct current (DC) and radio frequency (RF) signal from the photodetector can be simultaneously monitored with the voltmeter and the network analyzer, respectively.

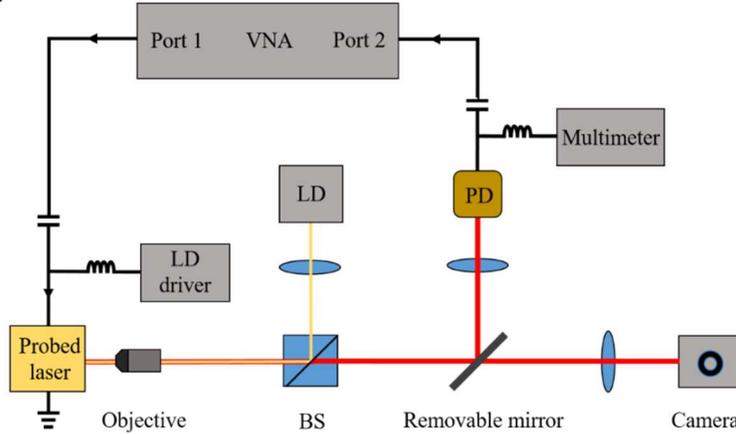

**Fig.5.** The schematic of modulation response measurement station.

In our measurements, we first apply a DC current to the straight bus waveguide to prevent the laser output from further attenuation. Since the device is symmetrically cleaved and one mode couples out from each arm, by leaving the other arm unpumped, optical feedback is prevented. With increasing injection current on the ring resonator, a laser emission is observed on the camera, as shown in Fig.6(a). The evolution of spectrum is displayed in Fig.7(a), showing a laser threshold current of around 27mA. When the laser is operated slightly above threshold, multiple longitudinal modes are observed, while single mode emission is obtained under higher pumping levels. The L-I and I-V curves of the two-ring system are shown in Fig.6(b), indicating that the laser threshold current is consistent with that obtained from the spectral evolution. The electrical resistance of the probed laser $R_{LD}$ is determined from the I-V curve as $R_{LD} = R_{total} - R_{load} = 5\,\Omega$, displaying a good ohmic contact at the metal-semiconductor boundary. The saturated output power is limited by confinement factor, heat, etc. In our design, the waveguides are etched below the MQW region in order to achieve a low bending loss, thus the surrounding dielectric layer (BCB) has a lower thermal conductivity than InP (0.3 W/mK for BCB [32] and 68 W/mK for InP [33]), leading to less effective thermal dissipation.

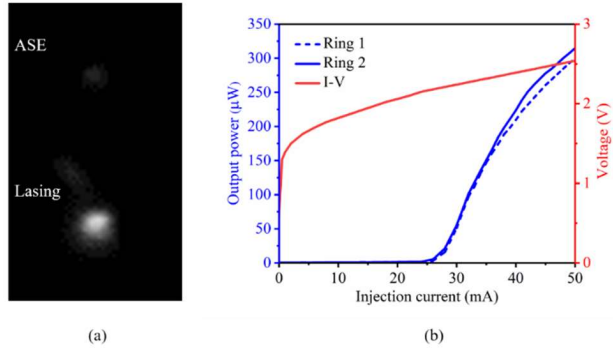

**Fig.6.** (a) Output beam imaged from the facet of the bus wasveguide; (b) LIV curves of microring lasers.

The $S_{21}$ of picoprobe-laser-detector system is directly measured by the network analyzer then normalized to the free running response in order to compensate the stopband of the network analyzer at low frequency. The frequency bandwidth of the picoprobe and the photodetector are 40 GHz and 15 GHz, respectively. As a result, any roll-off below 15 GHz can be attributed to the ring laser. For verifying the gain-loss contrast can tune the frequency response, ring 1 is biased beyond the threshold and ring 2 is pumped below the threshold. Therefore, $\delta$ is actually adjusted by the biased current of ring 2 $I_2$ ($\delta$ decreases with the increasing $I_2$). The frequency response of the ring laser with an injection current of $I_1$=35mA ($I_2$=0mA) is shown in Fig.7(b), and it exhibits a 3-dB bandwidth of 2.2 GHz (black curve). With an increase in the injection current of ring 2 while keeping $I_1$=35mA, the modulation bandwidth broadens to 3 GHz at $I_2$=22 mA (red curve) and to 3.6 GHz at $I_2$=23 mA (blue curve). The lasing spectrum for $I_2$=23 mA is shown in Fig.7(a), exhibiting no obvious change with respect to that of a single ring. Further increasing $I_2$ will not broaden the bandwidth but drop it down (orange curve). System RF response is shown in the Fig. 7(b) as a reference. During the measurement, the DC compoment of photon current is monitored in order to make sure that the average photon density of ring 1 remains constant. Therefore, By tuning the injection current of the lossy ring, a modulation bandwidth enhancement of 1.63 times was observed without increasing the photon density of ring 1.

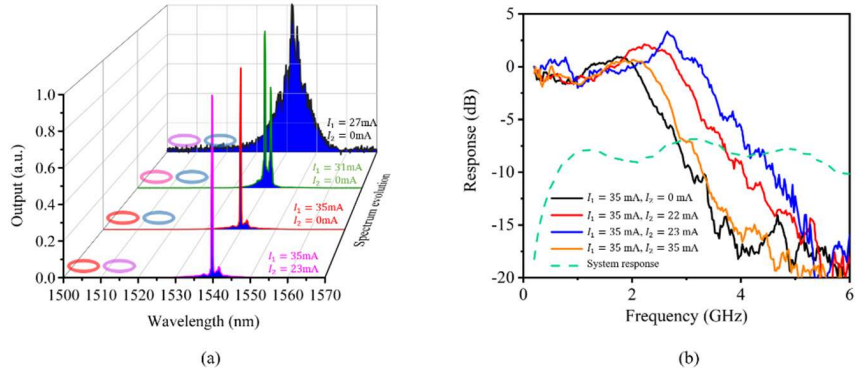

**Fig.7.** (a) Spectrum evolution; (b) modulation responses with different pumping ratios between two rings.

## 6. Conclusion

In conclusion, we have demonstrated gain-loss contrast (differential gain) as a new knob that can be used to increase the modulation bandwidth of semiconductor microring lasers. An electrically pumped III-V semiconductor laser system, comprised of two coupled deeply-etched ring resonators, was fabricated. It achieved a continuous wave lasing operation at room temperature with a threshold current of 27 mA. By increasing the injection current of the lossy ring while keeping the photon density of the gain ring invariant, an enhancement of modulation bandwidth by up to 1.63 times over that of the single ring has been observed. We believe this new paradigm could help pave the way for a new generation of directly modulated on-chip light sources. Future work will focus on improving the design and fabrication process for lowering the threshold, increasing the output power, and enhancing the modulation speed.

## Funding

## Disclosures

The authors declare no conflict of interest.

## Data availability

Data underlying the results presented in this paper are not publicly available at this time but may be obtained from the authors upon reasonable request.